\documentclass[10pt,letterpaper]{article}
\usepackage[top=1in,left=1.2in,right=1.2in, footskip=0.5in]{geometry}

\usepackage{amsmath,amssymb}

\usepackage{changepage}

\usepackage[utf8]{inputenc}

\usepackage{textcomp,marvosym}

\usepackage{nameref,hyperref}

\usepackage[right]{lineno}

\usepackage{microtype}
\DisableLigatures[f]{encoding = *, family = * }

\usepackage[table]{xcolor}
\usepackage{float}

\usepackage[aboveskip=1pt,labelfont=bf,labelsep=period,singlelinecheck=off]{caption}

\usepackage[
sortcites,
sorting=none,
url=false,
isbn=false,
doi=false,
eprint=false]{biblatex}
\bibliography{bib/biblio}
\AtEveryBibitem{
 \clearlist{language}
 \clearlist{editor}
 \clearfield{note}
}

\makeatletter
\renewcommand{\@biblabel}[1]{\quad#1.}
\makeatother

\usepackage{graphicx}

\usepackage{multicol}

\usepackage{calc}

\newcommand{\textover}[3][l]{%
 \makebox[\widthof{#3}][#1]{#2}%
 }

\usepackage{xcolor}
\usepackage{graphicx}
\usepackage{tcolorbox}
\tcbuselibrary{skins, theorems}
\usepackage{nameref}            

\definecolor{boxTitle}{HTML}{fff79a}
\definecolor{boxBackground}{HTML}{fffce0}
\definecolor{boxFrame}{HTML}{f1e2b8}

\tcbset{my box/.style={
    enhanced, fonttitle=\bfseries,
    colback=boxBackground, colframe=boxFrame,
    coltitle=black, colbacktitle=boxTitle,
    attach boxed title to top left={xshift=0.3cm,
                                    yshift*=-\tcboxedtitleheight/2},
    boxed title style={
    }
  }
}

\newtcolorbox[auto counter, number within=section]{infobox}[2][]{%
  my box, title={Infobox~\thetcbcounter: #2}, #1}

\newcommand{\hpc}{hPC}
\newcommand{\dec}{dendritic \hpc{}}

\title{
Dendritic predictive coding: \\
A theory of cortical computation with spiking neurons
}

\author{Fabian A. Mikulasch\textsuperscript{1,*}, Lucas Rudelt\textsuperscript{1,*}, Michael Wibral\textsuperscript{2}, Viola Priesemann\textsuperscript{1,3} \\
\small
\textbf{1} Max-Planck-Institute for Dynamics and Self-Organization, Göttingen, Germany
\\
\small
\textbf{2} Campus Institute for Dynamics of Biological Networks, Georg-August University, Göttingen, Germany
\\
\small
\textbf{3} Bernstein Center for Computational Neuroscience (BCCN), Göttingen, Germany
\\
\small
* These authors contributed equally
}
\date{}

\begin{document}

\maketitle

Top-down feedback in cortex is critical for guiding sensory processing, which has prominently been formalized in the theory of hierarchical predictive coding (hPC).
However, experimental evidence for error units, which are central to the theory, is inconclusive, and it remains unclear how hPC can be implemented with spiking neurons.
To address this, we connect hPC to existing work on efficient coding in balanced networks with lateral inhibition, and predictive computation at apical dendrites.
Together, this work points to an efficient implementation of hPC with spiking neurons, where prediction errors are computed not in separate units, but locally in dendritic compartments.
The implied model shows a remarkable correspondence to experimentally observed cortical connectivity patterns, plasticity and dynamics, and at the same time can explain hallmarks of predictive processing, such as mismatch responses, in cortex.
We thus propose dendritic predictive coding as one of the main organizational principles of cortex.

\section{Introduction}
A central feature of perception is that our internal expectations to a large degree shape how we perceive the world~\cite{de2018expectations}.
A long line of research aims to describe these expectation-guided computations in our brain by Bayesian inference, i.e. statistically optimal perception \cite{fiser2010statistically}, and subsequently could show that Bayesian inference can capture perception extraordinarily well~\cite{de2018expectations}. 
In light of this success, it has been proposed that the primary computation that is performed by the cortex is a hierarchically organized inference process, where cortical areas combine bottom-up sensory information and top-down expectations to find a consistent explanation of sensory data \cite{lee2003hierarchical,bastos_canonical_2012,aitchison_or_2017}. 

While the general idea of hierarchical inference in cortex found considerable experimental support \cite{de2018expectations,aitchison_or_2017}, it is less clear how exactly this inference could be implemented by cortical neurons. 
A popular theory that has been proposed to describe the neural substrate of inference in cortex is classical hierarchical predictive coding (\hpc{})~\cite{rao_predictive_1999,bastos_canonical_2012}. A central proposition of this theory is the existence of error units, which are thought to compare top-down predictions to bottom-up inputs, and guide neural computation and plasticity.
However, classical \hpc{} for the most part remains on the level of firing-rate dynamics of neural populations, and it has proven difficult to connect the theory to the properties of single neurons with spiking dynamics~\cite{kogo2015predictive,millidge_predictive_2021}.

With this perspective we point towards a different, emerging theory of biologically plausible hierarchical inference, which relies on the local membrane dynamics in neural dendrites. 
The core idea of this theory, which we will refer to as \dec{}, is to shift error computation from separate neural populations into the dendritic compartments of pyramidal cells. As we will discuss, this shift in perspective for the first time enables a biologically plausible implementation of \hpc{} with spiking neurons, and bears the potential to explain how cortical computations are organized from the micro-scale of individual dendrites to the macro-scale of cortical areas. 

In the first part of the perspective, we will outline the theory of dendritic \hpc{} and explain how it allows for inference with biologically plausible spiking dynamics and synaptic plasticity.
In the process, our goal is to combine several theoretical insights that have been gained over the last years into a coherent picture of cortical computation. A central contribution of this section therefore is to point out the intimate relations that exist between three previously mostly independent branches of research of theoretical neuroscience: Hierarchical predictive coding \cite{rao_predictive_1999}, efficient coding in balanced spiking networks \cite{deneve_efficient_2016} and the theory of neural sampling in probabilistic models \cite{buesing2011neural}.
In the larger, second part, we will discuss the biological plausibility of the theory, and explain how several experimental observations of hierarchical cortical computation fit into the picture. Ultimately, we will discuss the open challenges and most interesting avenues of theoretical and empirical research that are opened up by this this new perspective.

\section{Dendritic predictive coding in balanced spiking neural networks}

\subsection{Classical models of predictive coding}
\begin{figure}[t]
    \centering
    \makebox[\textwidth][c]{
    \includegraphics{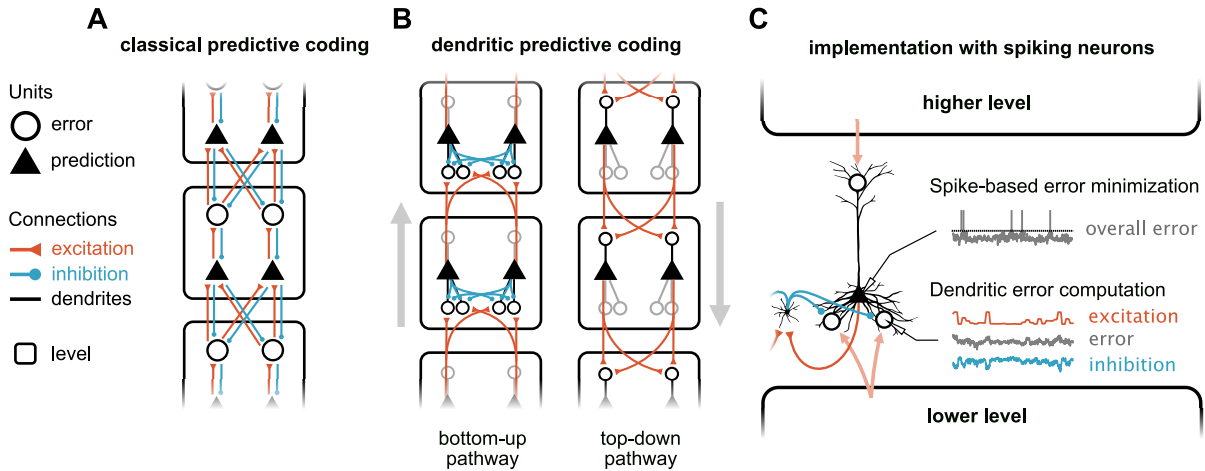}
    }
    \caption{\textbf{Implementation of predictive coding with dendritic error computation and spiking neurons.} \textbf{A}~Illustration of the classical model of hierarchical predictive coding (\hpc{}). Errors and predictions are computed in different neural populations within one level of the hierarchy. Errors are sent up the hierarchy, while predictions are sent downwards. \textbf{B}~In dendritic \hpc{}, prediction neurons implement the same function, but errors are computed in neural dendrites. Predictions are sent up the hierarchy to basal dendrites, where they are balanced by lateral connections to compute bottom-up prediction errors (left). At the same time, predictions are sent down the hierarchy to apical dendrites, where they try to predict somatic spiking and guide the inference process (right). The pathways are shown separately for better visibility. \textbf{C}~Dendritic \hpc{} can be implemented with spiking neurons. The errors that are computed in the dendritic membrane potentials are integrated at the soma to form an overall error signal of the neuron's encoding. A spike is emitted when the somatic error potential grows too large, and a spike would lead to a reduction in the overall error.}
    \label{fig:circuit_diagram}
\end{figure}

Hierarchical predictive coding (\hpc{}) describes the processing of sensory information as inference in a hierarchical model of sensory data (see Infobox~\ref{info:generative_models} for the basic notions of inference in generative models; see Infobox~\ref{info:error_computation} for mathematical details, which are not needed to understand the main text).
The central idea of \hpc{} is that activity of prediction units in one level of the hierarchy
\begin{itemize}
    \item[i)] should accurately predict sensory data, or the prediction unit activity in a lower level, and
    \item[ii)] should be consistent with top-down predictions generated by higher levels in the hierarchy.
\end{itemize}
\hpc{} tries to understand how these properties of neural activity can be ensured by neural dynamics on short timescales, and neural learning and plasticity on long timescales.
The theory predicts that to this end the prediction units in every level of the hierarchy need access to two types of errors:
\begin{itemize}
    \item[i)] Bottom-up errors, i.e. the mismatch between predictions generated within the level and the activity in lower levels. 
    \item[ii)] Top-down errors, i.e. the mismatch between top-down predictions and the activity within the level.
\end{itemize}
A neural implementation thus also requires a neural substrate for these errors. In classical \hpc{}~\cite{rao_predictive_1999}, the key innovation was to represent these errors in distinct populations of error units that compare top-down predictions with the activity within a level (Fig~\ref{fig:circuit_diagram}A). 
The elegance of this approach is that the same error units can mediate both, bottom-up errors to update prediction units in the next level, as well as top-down errors to neurons of the same level. Another central result of classical \hpc{} is that the learning rules that improve the hierarchical model take the form $[\text{error}\times\text{prediction}]$, which in the proposed architecture turns out to be classic Hebbian plasticity, i.e. the multiplication of pre- and postsynaptic activity.

\begin{figure}
\begin{infobox}[label={info:error_computation}]{Mathematical details of hierarchical predictive coding}
\begin{multicols}{2}
\small
\paragraph{Predictive coding with error units.} 
The goal in \hpc{} is to maximize the model log-likelihood~\cite{rao_predictive_1999} (for a detailed tutorial see~\cite{bogacz2017tutorial})
\begin{equation}
    \mathcal{L} = \sum_{i=1}^N \log p_{\theta}(\mathbf{r}^{i-1}|\mathbf{r}^{i}), \label{eq:model_likelihood}
\end{equation}
where $\theta$ are the model parameters, $\mathbf{r}^i$ is neural activity of a neural network at level $i$, and inputs are provided by the previous level $\mathbf{r}^{i-1}$.
This defines a hierarchy of processing stages that for example can be associated with different visual cortical areas (e.g. V1, V2, etc.), where $\mathbf{r}^{0}$ are visual inputs from LGN~\cite{rao_predictive_1999}. 
Typically, a linear model is assumed, where inputs are modelled according to 
\begin{equation}
    \mathbf{r}^{i-1} = \mathbf{D}^{i} \mathbf{r}^{i} + \mathbf{n}^{i-1},
\end{equation}
with decoding matrix $\mathbf{D}^i$ and Gaussian white noise $\mathbf{n}^{i-1}$ with zero mean and variance $\sigma_{i-1}^2$. 
With this model, for a single level $i$, the relevant contributions of the negative log-likelihood $- \mathcal{L}^i$ take the intuitive form of the square sum of coding errors for bottom-up inputs and errors of top-down predictions:
\begin{align}
\begin{split}
   &\text{bottom-up error:} \quad  \mathbf{e}^{i-1} = \mathbf{r}^{i-1} - \mathbf{D}^i \mathbf{r}^{i}, \\
   &\textover{top-down error:}{bottom-up error:}  \quad  \mathbf{e}^{i} = \mathbf{r}^{i} - \mathbf{D}^{i+1} \mathbf{r}^{i+1} ,
\end{split}
\end{align}
\begin{align}
   - \mathcal{L}^i = 
   \frac{1}{2\sigma_{i-1}^2} {\mathbf{e}^{i-1}}^T \mathbf{e}^{i-1}
   + 
   \frac{1}{2\sigma_{i}^2} {\mathbf{e}^{i}}^T \mathbf{e}^{i}. \label{eq:model_likelihood_gauss}
\end{align}
The goal is then to minimize the sum of coding errors on a fast timescale $\tau_r$ via neural dynamics $\frac{d}{dt}\mathbf{r}^{i}$, and with a slow learning rate $\eta_D$ via neural plasticity on the weights $\mathbf{D}^i$, by performing gradient ascent on $\mathcal{L}$:
\begin{align}
    &\text{dynamics:} \quad \tau_r \frac{d}{dt}\mathbf{r}^{i} =  \frac{1}{\sigma_{i-1}^2} {\mathbf{D}^i}^T \mathbf{e}^{i-1} - \frac{1}{\sigma_{i}^2} \mathbf{e}^i \label{eq:inference_dynamics}\\
    &\text{plasticity:} \quad \eta_D^{-1} \frac{d}{dt}\mathbf{D}^i  =  \frac{1}{\sigma_{i-1}^2} \mathbf{e}^{i-1} {\mathbf{r}^i}^T.
\end{align}
To yield a neural implementation, the key innovation in classical \hpc{} was to represent prediction errors within a distinct neural population of error units, that integrate inputs of prediction units within the same level, and subtract top-down predictions according to
\begin{equation}
     \tau_e \frac{d}{dt}\mathbf{e}^{i} = -  \mathbf{e}^{i} + \mathbf{r}^{i} -  \mathbf{D}^{i+1} \mathbf{r}^{i+1}, \label{eq:error_dynamics}
\end{equation}
where decoding weights $\mathbf{D}^i$ now correspond directly to weights of neural connections \cite{bogacz2017tutorial}.
Together with the dynamics of prediction units, this results in the hierarchical neural circuit shown in Fig~\ref{fig:circuit_diagram}A.

\paragraph{Predictive coding with dendritic errors.}
In \dec{}, the computation of errors in Eq~(\ref{eq:error_dynamics}) is accomplished by the leaky voltage dynamics of dendritic compartments~\cite{boerlin2011spike, deneve_efficient_2016, mikulasch_2021}. To this end, for each prediction neuron $j$ one introduces basal dendritic compartments
$b^i_{jk}\approx D^{i}_{kj}e^{i-1}_k$, which are each innervated by a single synapse of a prediction neuron $k$ of the previous level, as well as an apical compartment $a^i_j\approx -e^i_j$ that is innervated by prediction neurons of a higher level (Fig~\ref{fig:circuit_diagram}B). The error computation is then performed by voltage dynamics according to
\begin{align}
    \tau_b\frac{d}{dt} b^i_{jk} &= -b^i_{jk} + D^{i}_{kj} r^{i-1}_k - \sum_l W^{i}_{jkl} r^{i}_l,\label{eq:basal_dendrite_dynamics}\\
    \tau_a \frac{d}{dt} a^i_j & = -a^i_j - r^i_j + \sum_l D^{i+1}_{jl} r^{i+1}_l ,\label{eq:apical_dendrite_dynamics}
\end{align}
where bottom-up inputs are balanced with lateral connections $W^{i}_{jkl}$ (connection of neuron $r^i_l$ to the $k$th dendritic compartment of neuron $r^i_j$), and top-down predictions are matched by the neurons own predictions $r^i_j$.
The latter could be implemented via the backpropagating action potential~\cite{urbanczik2014learning}, solving the one-to-one connections problem of classical \hpc{}~\cite{millidge2020relaxing}.
To compute bottom-up errors, lateral weights have to be chosen as $W^{i}_{jkl} = D^{i}_{kj}D^{i}_{lj}$. This can be achieved if lateral plasticity enforces a tight balance~\cite{mikulasch_2021}
\begin{equation}
    \eta_W^{-1} \frac{d}{dt} W^i_{jkl} = \frac{1}{\sigma_{i-1}^2} b^{i}_{jk} r^i_l.   
\end{equation}

The dynamics of prediction neurons are then simply driven by the dendritic error potentials
\begin{equation}
    \tau_r \frac{d}{dt} r^{i}_j = \frac{1}{\sigma_{i-1}^2} \sum_k b^i_{jk} + \frac{1}{\sigma_{i}^2} a_j^i,
    \label{eq:dendritic_hpc_dynamics}
\end{equation}
and weights for bottom-up and top-down inputs can be learned with voltage-dependent rules
\begin{align}
     \eta_D^{-1}\frac{d}{dt} D^i_{kj} &= \frac{1}{\sigma_{i-1}^2} \frac{1}{D^i_{kj}} b^{i}_{jk} r^i_j, \label{eq:basal_syn_plasticity}\\
     \eta_D^{-1} \frac{d}{dt} D^{i+1}_{jl} &= - \frac{1}{\sigma_{i}^2} a^{i}_{j} r^{i+1}_l.
\end{align}
Here, correct learning requires the cooperation of lateral and bottom-up weights, which in classical \hpc{} is known as the weight transport problem~\cite{millidge2020relaxing,alonso_tightening_2021}. For \dec{} this problem has so far been addressed in the single-level case~\cite{mikulasch_2021}.

Together, these equations yield an equivalent formulation of \hpc{} for both, learning and inference, where prediction errors are computed in tightly balanced dendritic compartments. 
\end{multicols}
\end{infobox}
\end{figure}

\subsection{A functionally equivalent formulation of predictive coding with local dendritic error computation}

Although the idea of error units is undeniably elegant, it is not the only way to compute prediction errors in a neural circuit.
Mathematically equivalent, one may understand error computation as an operation that is performed in the voltage dynamics of individual dendritic compartments (see Infobox~\ref{info:error_computation} for mathematical details)---and, thus, without specialized error units. In this reinterpretation of \hpc{} every neuron will represent the two types of errors we discussed before in different sections of its dendritic tree (Fig~\ref{fig:circuit_diagram}B):
\begin{itemize}
    \item[i)] Bottom-up errors in basal dendritic compartments, where input from lower-level cortical areas is integrated~\cite{harris2015neocortical}.
    \item[ii)] The top-down prediction error (for the neuron's own activity) in an apical compartment, where higher-level cortical feedback arrives~\cite{harris2015neocortical}.
\end{itemize}
Besides the absence of error neurons, two additional central differences arise between the architectures of classical \hpc{}, and this \dec{}.
First, in \dec{} both bottom-up and top-down signals are predictions. The possibility to implement \hpc{} in such a manner (by redrawing hierarchical-level boundaries) has previously been discussed by Spratling~\cite{spratling_predictive_2008}. 
Second, and more importantly, while prediction units in classical \hpc{} inhibit each other through error units, prediction neurons in \dec{} directly compete through lateral inhibition on basal dendrites. 
Such networks with strong lateral inhibition (or similarly, winner-take-all-like dynamics) have a long tradition in theoretical neuroscience, as models for the sparse and efficient encoding of sensory data~\cite{foldiak1990forming, olshausen_emergence_1996, bourdoukan2012learning, mikulasch_2021} and as divisive normalization models of cortical computation~\cite{carandini2012normalization,burg2021learning} (see also \cite{douglas_neuronal_2004}). 
Dendritic \hpc{} is closely related to these models with lateral inhibition, except that these models so far did not consider how exactly top-down connections could guide neural computations with predictions.
In a more general context it has been proposed that top-down connections could provide these predictions by targeting apical dendrites~\cite{urbanczik2014learning,larkum2013cellular,gillon2021learning} (a similar architecture has also been used to implement backprop in a cortical microcircuit~\cite{sacramento2018dendritic}, see Infobox~\ref{info:dendritic_errors_hpc_versus_backprop} for a comparison).
With \dec{}, we propose that these ideas of hierarchical predictive coding and lateral competition can be combined to form a consistent theory of cortical computation. 

Since error computation in \dec{} takes place in the voltage dynamics of basal and apical dendritic compartments, in the model these local potentials play an important role for synaptic plasticity. 
For basal dendrites, \dec{} predicts that lateral connections between neurons should establish a tight balance locally, and thereby compute the errors for bottom-up inputs~\cite{mikulasch_2021}.
The intuitive explanation for this computation is that in a tightly balanced state, every input that can be predicted from other neurons is effectively canceled, and the remaining unpredictable input constitutes the prediction error~\cite{deneve_efficient_2016}. 
These errors can then be exploited by another voltage-dependent rule for bottom-up connections, in order to find an optimal encoding of inputs~\cite{mikulasch_2021}. 
This second rule is in general Hebbian, but strong local inhibition during a postsynaptic spike would signal an over-prediction of a specific input, which consequently should lead to the local depression of bottom-up synapses.
For apical dendrites it has been proposed that the error computation relies on the back-propagating action potential~\cite{urbanczik2014learning}. 
In this theory of apical learning, plasticity of top-down connections is Hebbian as well, but should be depressing for a depolarization of the apical dendritic potential in the absence of postsynaptic activity. 
By employing these voltage-dependent plasticity rules, \dec{} implements the same learning algorithm as classical \hpc{}, but in prediction neurons with dendritic error computation (Infobox~\ref{info:error_computation}).

\begin{figure}
\begin{infobox}[label={info:dendritic_errors_hpc_versus_backprop}]{Relation to dendritic microcircuits for error backpropagation}
\begin{multicols}{2}
\small

Closely related to \dec{}, recent work has explored the role dendrites could play in hierarchical computation and learning in cortex \cite{sacramento2018dendritic,richards_dendritic_2019,whittington2019theories}. Most relevant, Sacramento et al. have proposed that a predictive coding-like dendritic microcircuit can implement the backpropagation algorithm~\cite{sacramento2018dendritic}. While sharing many concepts, \dec{} and the dendritic error model by Sacramento et al. pursue different goals, which ultimately lead to different architectures (Fig~A). 

The prime difference is that in \dec{} neurons compute prediction errors via a balance on basal dendrites, while in \cite{sacramento2018dendritic} neurons compute errors of the network's output to a target via a balance on apical dendrites. 
In support of the \dec{} model, the interneuron architecture in cortex is well adapted to provide fast lateral inhibition to basal dendrites, to rapidly compute prediction errors online during sensory processing; apical inhibition in comparison is rather slow, meaning that apical error computation would require a slow equilibration process (interneurons are discussed more in-depth in section \ref{sec:interneurons}).
Furthermore, recent experimental observations seem to be at odds with the idea that a balance of inputs computes an error at apical dendrites, and are more consistent with a predictive role of apical activity~\cite{gillon2021learning,chen2015subtype}.
However, these particular results of course do not rule out that cortical networks could make use of both proposed mechanisms in different modes of operation or different neural populations, e.g. depending on the task at hand.

\vspace{8mm}
\begin{minipage}[t]{0.485\textwidth}
\begin{center}
\includegraphics{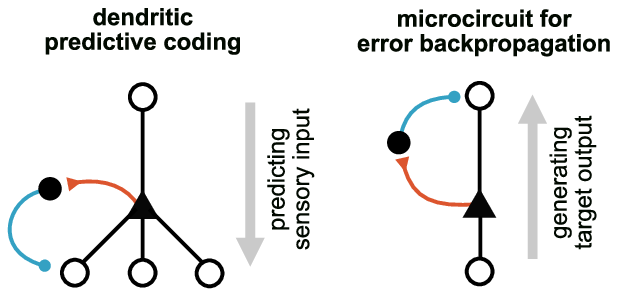}
\end{center}
\footnotesize\textit{Fig~A. 
(Left) In \dec{} the goal is to generate predictions of bottom-up sensory inputs, and the corresponding prediction errors are computed via inhibition to basal dendrites.
(Right) In \cite{sacramento2018dendritic} the goal is to generate a target output at the highest level (e.g. a label). To this end an 'inverted' model of \hpc{} is employed \cite{whittington2019theories}, where inhibition at the apical dendrite is used to compute the backpropagated error of the output.} 
\end{minipage}

\end{multicols}
\end{infobox}
\end{figure}

\subsection{Dendritic errors enable an efficient implementation of hierarchical predictive coding with spiking neurons}

Dendritic errors do not only yield an equivalent formulation of hierarchical predictive coding, they also enable inference with spiking neurons. 
Here, the inferred variables have to be efficiently represented by spikes, which is possible if spikes are only fired if they reduce the overall prediction error~\cite{boerlin2011spike,deneve_efficient_2016,kadmon2020predictive, yoon2016lif} (see Infobox~\ref{info:spiking_implementation} for the mathematical details of \dec{} with spiking neurons). 
As we elaborated above, this prediction error can be computed by dendritic and somatic voltage dynamics if lateral (inhibitory) connections enforce a tight balance between excitatory and inhibitory currents~\cite{deneve_efficient_2016}, where predictions track bottom-up input on fast timescales. 
In this case, an efficient spike encoding can be found with a simple threshold mechanism on the membrane potential that generates a spike when the error potential grows too large~\cite{boerlin2011spike}, similar to the generation of action potentials in biological neurons (Fig~\ref{fig:circuit_diagram}C). 
Predictive coding thus serves a dual purpose here, by enabling both inference in a hierarchical model, and an efficient spike encoding of dynamical variables.

A central role in this inference scheme with spikes is played by noise in the neural dynamics, for two reasons.
First, noise enables an efficient spiking code in the face of transmission delays.
With deterministic neurons, even a small delay of inhibition can lead to erratic network behavior, since inhibition will often arrive too late to prevent synchronous spiking of large parts of a population~\cite{buxo2020poisson}.
Noise relaxes this constraint on the speed of feedback, since it effectively decouples and desynchronizes neural spiking~\cite{kadmon2020predictive, buxo2020poisson, chalk2016neural}.
Second, noise in spiking neural networks enables neural sampling~\cite{savin2014spatio,buesing2011neural,aitchison2016hamiltonian}. Here, the idea is that neural activity samples possible predictions according to their likelihood, instead of computing a single best estimate as in classical \hpc{} (Infobox~\ref{info:spiking_implementation}). 
Neural sampling therefore is a principled way to represent uncertainty in inference via neural activity, and has for example been used to explain variability in neural responses \cite{orban2016neural}, and the origin of multistability in perception~\cite{gershman2012multistability}.
Recent models show that neural sampling and efficient spike coding with tight excitation-inhibition balance can be combined in a single model with dendritic error computation~\cite{savin2014spatio,aitchison2016hamiltonian,mikulasch_2021}, relating these concepts to the proposed model of \dec{} (Infobox~\ref{info:spiking_implementation}).

In addition to neural inference, dendritic errors also enable \emph{learning} in populations of spiking neurons.
This is not straightforward, since the switch from rate-based to spike-based models typically requires a modification of the learning algorithms. 
For example, 
when using populations of spiking error units as in classical \hpc{}, prediction errors will have to be approximated by discrete events of spikes, and further approximations are necessary to represent both positive and negative values with spikes or non-negative firing rates~\cite{alonso_tightening_2021}.
To resolve this, it was proposed that errors are represented by deviations relative to a baseline firing rate~\cite{alonso_tightening_2021}, but this would require high firing rates and therefore seems implausible considering the low firing rates in neocortex~\cite{keller_predictive_2018}. An alternative is to represent positive and negative errors in separate populations~\cite{rao_predictive_1999,keller_predictive_2018}, but it is unclear how in this case biological plasticity can recombine the positive and negative parts, which are both required for the learning of single synapses. 
Due to these difficulties, to the best of our knowledge, as of yet no complete implementation of \hpc{} has been proposed that uses spiking error units (see also~\cite{millidge_predictive_2021} for a recent and extensive review).
In contrast, in \dec{} the same learning algorithm as for rate-based units can be straightforwardly applied to spiking neurons.
The reason is that dendritic membrane potentials remain continuous quantities that can attain positive and negative values relative to some reference value, despite the spiking nature of neural responses.
Therefore, these dendritic potentials can easily represent the prediction errors that are required for the learning of bottom-up and top-down connections (Infobox~\ref{info:error_computation}), which has been successfully applied in~\cite{mikulasch_2021, urbanczik2014learning}.

\begin{figure}
\begin{infobox}[label={info:spiking_implementation}]{Mathematical details of dendritic predictive coding with spikes}
\begin{multicols}{2}
\small

\paragraph{Spike-based predictions of sensory data.}
In spiking neural networks, predictions are not based on scalar firing rates that attain some steady-state, but are dynamic quantities that change with each event of a spike. A popular choice to mathematically formalize the prediction generated by a spike at time $t_{sp}$ is via spike traces $\kappa(t, t_{sp})= \exp(-(t-t_{sp})/\tau)$ that decay exponentially with some time constant $\tau$ \cite{boerlin2011spike,mikulasch_2021}. Predictions of a neuron $j$ on level $i$ then change upon a spike according to $r_j^i(t) \rightarrow r_j^i(t) + \kappa(t, t_{\text{sp}})$. This approximately corresponds to the way spikes are integrated in the membranes of postsynaptic neurons, and is in line with the intuition that spikes are more predictive for inputs close to spike time than for inputs further in the future. With these predictions $r_j^i(t)$, the same formalism as before can be used to compute the instantaneous log-likelihood (Infobox~\ref{info:error_computation}):
\begin{equation}
    \mathcal{L}(t)= \sum_{i=1}^N \log p_{\theta}(\mathbf{r}^{i-1}(t)|\mathbf{r}^{i}(t)).
\end{equation}
However, inference can no longer be implemented by simple gradient descent. Because spikes are predicting future data, finding an efficient spiking implementation of predictive coding becomes itself a predictive coding problem.

\vspace{5mm}
\paragraph{Efficient spiking implementation of predictive coding with dendritic errors.}
One straightforward approach to implement inference with spikes is to deterministically fire a spike at time $t$ if it instantly improves bottom-up and top-down errors, i.e. the log-likelihood $\mathcal{L}(t)$~\cite{boerlin2011spike}:
\begin{equation}
\begin{aligned}
  &\mathcal{L}(t|\text{neuron }j\text{ spikes at time }t) \\ 
  > &\mathcal{L}(t|\text{no spike at time }t).
  \end{aligned}\label{eq:spiking_condition}
\end{equation}
This can be seen as a discrete implementation of gradient ascent to find the maximum a posteriori (MAP) estimate for predictions $r_j^i$. The resulting spike code can be considered efficient, since a spike is only fired if it contributes to improving the likelihood, otherwise neurons stay silent. As it turns out, this algorithm can be implemented by a simple spiking threshold mechanism together with the principle of tight balance~\cite{boerlin2011spike}, where a neuron fires a spike if its membrane potential $u^i_j(t)$ surpasses a firing threshold $T_j$, i.e.
\begin{equation}
   u_j^i(t) = \frac{1}{\sigma_{i-1}^2} \sum_k b^i_{jk} + \frac{1}{\sigma_{i}^2} a_j^i > T_j.\label{eq:threshold_condition}
\end{equation}
Here, $b^i_{jk}(t)$ are the balanced dendritic potentials of basal dendrites and $a^i_j(t)$ the potential of the apical dendrite as in Infobox~\ref{info:error_computation}, which represent bottom-up or top-down errors, respectively (Fig~A). The somatic membrane potential $u^i_j(t)$ thus represents a projection of the total error, enabling neurons to efficiently maximize the model likelihood~(Fig~\ref{fig:circuit_diagram}C).

\vspace{5mm}
\begin{minipage}[t]{0.485\textwidth}
\begin{center}
\includegraphics{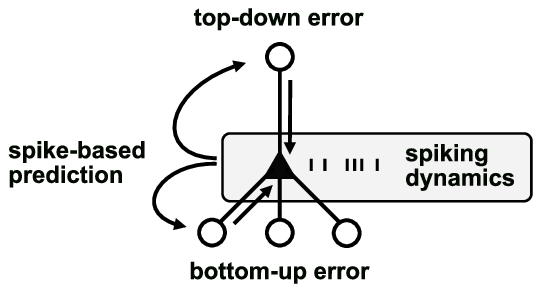}
\end{center}
\footnotesize\textit{Fig~A.} An efficient spiking implementation of inference can be found if spiking dynamics consist of a tight feedback loop between discrete spiking and continuous prediction errors. This tight feedback loop can be implemented with lateral connections between neurons and the local voltage dynamics that compute errors in dendritic compartments. 
\end{minipage}

\paragraph{Predictive coding with neural sampling.}
Instead of finding an instantaneous MAP estimate, a more general approach to inference with spikes is to sample a collection of (binary) spikes $\mathbf{S}_{0..T}=\{\mathbf{s}(t) | t \in \{0..T\}\}$ from the posterior distribution of the generative model $\mathbf{S}_{0..T} \sim p_\theta(\mathbf{S}_{0..T}|\mathbf{r}^0_{0..T})$~\cite{mikulasch_2021,kappel_stdp_2014}. A central problem of this approach is that this distribution cannot be computed at any time $t$, since information about future inputs is missing and computing the normalization of the distribution is intractable~\cite{mikulasch_2021,kappel_stdp_2014}. Yet, using simple approximations, one can show that approximate sampling can be implemented with the same membrane potentials $u_j^i(t)$ and threshold $T_j$ as before (up to a constant factor), and a soft spiking threshold mechanism
\begin{equation}
   p(\text{neuron } j \text{ spikes at time }t) = \text{sig}(u_j^i(t) - T_j), \label{eq:sampling_mechanism}
\end{equation}
where $\text{sig}(x)=1/(1+\exp(-x))$ is the logistic function~\cite{mikulasch_2021,savin2014spatio}. 
This model is a special case of the spike response model with escape noise~\cite{gerstner2014neuronal}, and can be implemented by a standard leaky integrate and fire neuron with a noisy membrane potential. 
Equations~(\ref{eq:dendritic_hpc_dynamics}), (\ref{eq:threshold_condition}) and (\ref{eq:sampling_mechanism}) highlight the intimate relation that exists between the theories of \hpc{}, efficient coding with spikes and neural sampling.
\end{multicols}
\end{infobox}
\end{figure}

\section{Is dendritic predictive coding biologically plausible?}
\label{sec:plausibility}

In the previous section we have introduced the two main assumptions of \dec{}, which are that i) cortex implements inference in a hierarchical probabilistic model, and ii) errors of the resulting predictions are computed in the local voltage dynamics of basal and apical dendrites.
The implications of the first assumption have been discussed at length in the context of classical \hpc{}, and were found to align well with experimental observations~\cite{aitchison_or_2017,heilbron_2018,walsh2020evaluating}.
From the second assumption, and the combination of the two assumptions, additional requirements on neural lateral connectivity and plasticity arise, which we will discuss in the following.
As we will see, these requirements are remarkably in line with properties of cortical pyramidal neurons and inhibitory connectivity in cortex, which renders \dec{} a promising theory to explain their functions in cortical computations.

\subsection{Dendritic error computation and synaptic plasticity in pyramidal neurons}

To compute errors in basal dendrites, a tight and local excitation-inhibition (E-I) balance is required. 
That excitatory and inhibitory currents are balanced in cortical neurons is a well-established result~\cite{deneve_efficient_2016}, and it has been found in several instances that inhibition and excitation are tightly correlated, with inhibition trailing excitation by few milliseconds~\cite{deneve_efficient_2016,wehr2003balanced,okun2008instantaneous}.
This tight balance leaves neurons only with a brief window of opportunity for spiking, which effectively decorrelates neural responses to inputs and thereby ensures an efficient neural code~\cite{brendel2020learning}. 
A tight E-I balance can therefore explain the origin of the irregular spiking patterns of neurons that have been observed throughout cortex \cite{deneve_efficient_2016}. 
Although models with a tight balance can reproduce irregular firing on the single neuron level, incorporating realistic synaptic transmission delays in these models can lead to oscillations on the population level~\cite{kadmon2020predictive}. Oscillations in cortical activity (especially in the Gamma frequency band that dominates in layer 2/3~\cite{bastos_canonical_2012}) have therefore been discussed as signatures of efficient coding in balanced networks~\cite{chalk2016neural} (see also \cite{aitchison2016hamiltonian}). 
Thus, a tight E-I balance as assumed by \dec{} is not only observed experimentally at the neuron level, but also reproduces more global firing statistics of cortical activity.

Consistent with \dec{}, this balance has later been found to also extend to individual neural dendrites \cite{liu_local_2004,iascone2020whole,hennequin2017inhibitory}. Crucially, this local balance can be observed down to the scale of (at least) single dendritic branches \cite{liu_local_2004}, since the attenuation of dendritic currents prevents that inhibitory postsynaptic potentials spread into other dendritic compartments and influence the E/I balance there \cite{spruston2016principles,mullner2015precision}.
Experiments could also show that this local balance is maintained through localized synaptic plasticity, which reestablishes the balance after a perturbation, and coordinates excitatory and inhibitory plasticity locally~\cite{liu_local_2004,field2020heterosynaptic,chen2015subtype,hu2019endocannabinoid,bourne_coordination_2011,d2015inhibitory}. 
Overall, these findings are compatible with the idea that a local balance computes prediction errors for specific synaptic contacts at basal dendrites.

Assuming that local membrane potentials indeed reflect an error signal, \dec{} predicts that these potentials should influence synaptic plasticity locally~\cite{mikulasch_2021,urbanczik2014learning}. Such voltage-dependent plasticity (VDP) has been observed in a range of experiments \cite{artola1990different,lisman2005postsynaptic,lisman2010questions}, and is thought to be mainly mediated by the local calcium concentration, which follows the local membrane potential and modulates synaptic plasticity \cite{mullner2015precision,higley2014localized,augustine2003local}. Based on these observations, VDP rules have been proposed that can reproduce several experiments of spike-timing dependent plasticity in a unified picture \cite{shouval2002unified,clopath2010voltage}. A consequence of locally organized VDP is that inhibition strongly modulates synaptic plasticity in a very localized manner \cite{meredith2003maturation,hayama2013gaba,wang2014inhibitory,hennequin2017inhibitory,herstel2021network}; Hayama et al \cite{hayama2013gaba} for example found, that this modulation spreads less than 15 micrometers on dendrites. The local E/I balance therefore is a central determinant of synaptic plasticity, which---consistently with \dec{}---suggests that neurons purposefully incorporate it into neural learning. 

\begin{figure}[t]
    \centering
    \includegraphics{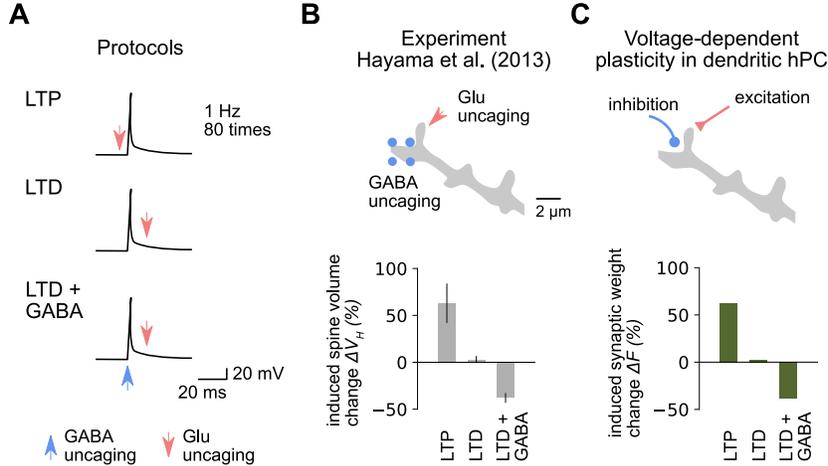}
     \caption{\textbf{Illustration of the effect of local inhibition on synaptic plasticity in experiment and dendritic hPC.}
\textbf{A}~In experiment, plasticity is induced in proximal dendrites with long-term potentiation (LTP) protocols or long-term depression (LTD) protocols with or without GABA uncaging (i.e. inhibition)~\cite{hayama2013gaba}. \textbf{B}~With GABA uncaging one observes a shrinkage of dendritic spine volume of nearby ($< 15\,\mu\text{m}$ distance) excitatory synapses, without GABA uncaging spine shrinkage is absent (values plotted as reported in~\cite{hayama2013gaba}). The relative change in spine volume serves as a proxy for synaptic weight change. 
\textbf{C}~The observed effect of local inhibition on synaptic plasticity is consistent with the voltage-dependent plasticity rule for bottom-up connections (Eq~(\ref{eq:basal_syn_plasticity}) in Infobox~\ref{info:error_computation}). 
The main aim here is to demonstrate the qualitative agreement between experiment and \dec{}. Yet, also a quantitative agreement exists if elicited postsynaptic potentials in the model are set to the typical EPSPs ($\approx 70\,\text{pA}$) and IPSPs ($\approx -45\,\text{pA}$) that are induced by glutamate and GABA uncaging, respectively~(Supplementary Fig~1 in~\cite{hayama2013gaba}), and assuming that temporal overlap between the backpropagating action potential and the induced EPSPs (LTP protocol) and IPSPs (LTD+GABA protocol) are the same. Learning rate and initial synaptic weight were tuned to match the experimental outcome for LTP and LTD. See \cite{github_plasticity} for further details and simulation code.}
    \label{fig:plasticity}
\end{figure}

Are the VDP rules that can be derived from \dec{} consistent with these experimentally observed VDP rules? A distinction has to be made here between VDP rules in basal dendrites, which should enable the learning of neural representations \cite{mikulasch_2021}, and VDP in apical dendrites, which should enable the prediction of somatic spiking \cite{urbanczik2014learning}. For representation learning in basal dendrites, we have argued that previously proposed VPD rules \cite{shouval2002unified,clopath2010voltage} can be reconciled with the VDP rules derived from \dec{} \cite{mikulasch_2021}. One prediction of these derived VDP rules is that strong local inhibition should promote the depression of excitatory synapses, an effect that has been observed in proximal dendrites of hippocampal pyramidal neurons \cite{hayama2013gaba} (Fig~\ref{fig:plasticity}; see also \cite{steele1999inhibitory}). This is interesting, since the selective depression of excitatory synapses in face of inhibition leads to a lesser E/I balance, contrary to the locally balancing plasticity discussed above. Yet, the interplay of such unbalancing (for bottom-up connections) and balancing (for lateral inhibition) VDP rules is perfectly consistent with \dec{}, and enables optimal representation learning \cite{mikulasch_2021}. For the learning of apical connections an explicit correspondence to experimental VDP still has to be proposed. Experiments show that synaptic plasticity close to and far from the soma behaves vastly different \cite{sjostrom2006cooperative,letzkus2006learning,gillon2021learning}, which could support the different requirements for basal and apical synaptic plasticity in \dec{}. 
While more experimental and theoretical work is needed to clarify the connections between \dec{} and experimental VDP, these results suggest that cortical pyramidal neurons in principle are suited to implement the learning algorithm proposed by \dec{}.

\subsection{A diversity of inhibitory interneurons is required for dendritic predictive coding}\label{sec:interneurons}           
Since pyramidal neurons in general only excite other cells, a diverse pool of inhibitory interneurons is required to implement the \dec{} model. Diverse, because inhibition has to act with different purposes on different parts of the dendritic tree of pyramidal neurons: 
First, inhibitory interneurons are needed that balance bottom-up inputs to basal dendrites via lateral connections, to compute bottom-up errors and coordinate neural responses~\cite{brendel2020learning}. 
Second, inhibitory interneurons are also needed for bottom-up inhibition of basal dendrites, which might require another type of interneuron since bottom-up inhibition relies on different plasticity rules than lateral inhibition~\cite{mikulasch_2021}.
Last, even more types of inhibitory interneurons would be required to mediate top-down inhibition, because inhibition alone at apical dendrites has a very weak effect at the soma, meaning that apical feedback is ineffective at suppressing somatic spiking \cite{spruston2016principles}.
To properly learn top-down inhibitory connections it therefore might be necessary to employ more complex interneuron circuits with disinhibition mechanisms, as suggested for example by models for the learning of prediction mismatch responses~\cite{hertag2020learning}. 

These are strong requirements, but indeed, cortex exhibits a wide array of inhibitory interneurons, some of which resemble the suggested connectivity patterns~\cite{kubota2014untangling,tremblay2016gabaergic}. One of the defining network motifs of cortex for example is implemented by parvalbumin-expressing (PV) interneurons, which mainly provide broad feedback inhibition within populations of pyramidal neurons~\cite{avermann2012microcircuits,petersen2013synaptic,tremblay2016gabaergic,douglas_neuronal_2004}. PV positive, fast-spiking basket cells alone make up around 40-50\% of all interneurons~\cite{kubota2014untangling}, and are especially adapted to tightly control pyramidal neuron spiking and the cortical E-I balance via very fast inhibition to somata and basal dendrites~\cite{tremblay2016gabaergic,ferguson2018pv}. Here, inhibitory interneurons can very specifically cancel thalamo-cortical (bottom-up) excitatory synapses~\cite{kubota2007neocortical}. This motif directly corresponds to the balancing lateral inhibition to basal dendrites that is central to \dec{}. PV interneurons also seem to be responsible for the Gamma oscillations that arise in \dec{}~\cite{ferguson2018pv}, further strengthening this link.
Somatostatin-expressing (SST) interneurons in contrast mainly inhibit apical dendrites and are typically innervated by facilitating excitatory synapses~\cite{tremblay2016gabaergic}, which enables them to more slowly control the influence of top-down inputs. 
A wealth of recent studies showed that these diverse inhibitory interneurons are crucial for cortical networks to respond with great stability to sensory inputs, and to incorporate feedback from other cortical areas into their computations \cite{herstel2021network}. 
Therefore, while the mapping of \dec{} to cortical interneuron connectivity will have to be made more explicit, the rich pool of cortical interneurons can certainly support the computations proposed by \dec{}.

\section{How can error responses arise in prediction neurons?}

One of the central features of classical \hpc{} is its ability to explain a variety of experimental observations through the concept of error neurons. 
Quite directly, error neurons can give a unified account of mismatch responses in cortex, which are neural responses that signal (or seem to signal) the mismatch between an internal model and sensory data, and would not be expected in a naive feedforward account of sensory processing~\cite{walsh2020evaluating}. Error neurons have also been argued to explain other observations that are seemingly unrelated to error computation~\cite{walsh2020evaluating}, such as extra-classical receptive field effects in visual cortex~\cite{rao_predictive_1999}.
An important question for \dec{} is, therefore, if these experimental observations can also be explained in a model without error neurons. How some of these effects can occur in models without error units, such as omission responses or expectation suppression, has been discussed before~\cite{walsh2020evaluating,aitchison_or_2017}. In the following we will focus on the less discussed effects of mismatch responses in visual cortex~\cite{zmarz2016mismatch} and top-down modulated extra-classical receptive field effects~\cite{rao_predictive_1999}, and outline how these types of responses can emerge in the prediction neurons of \dec{}.

\begin{figure}
    \centering
    \includegraphics{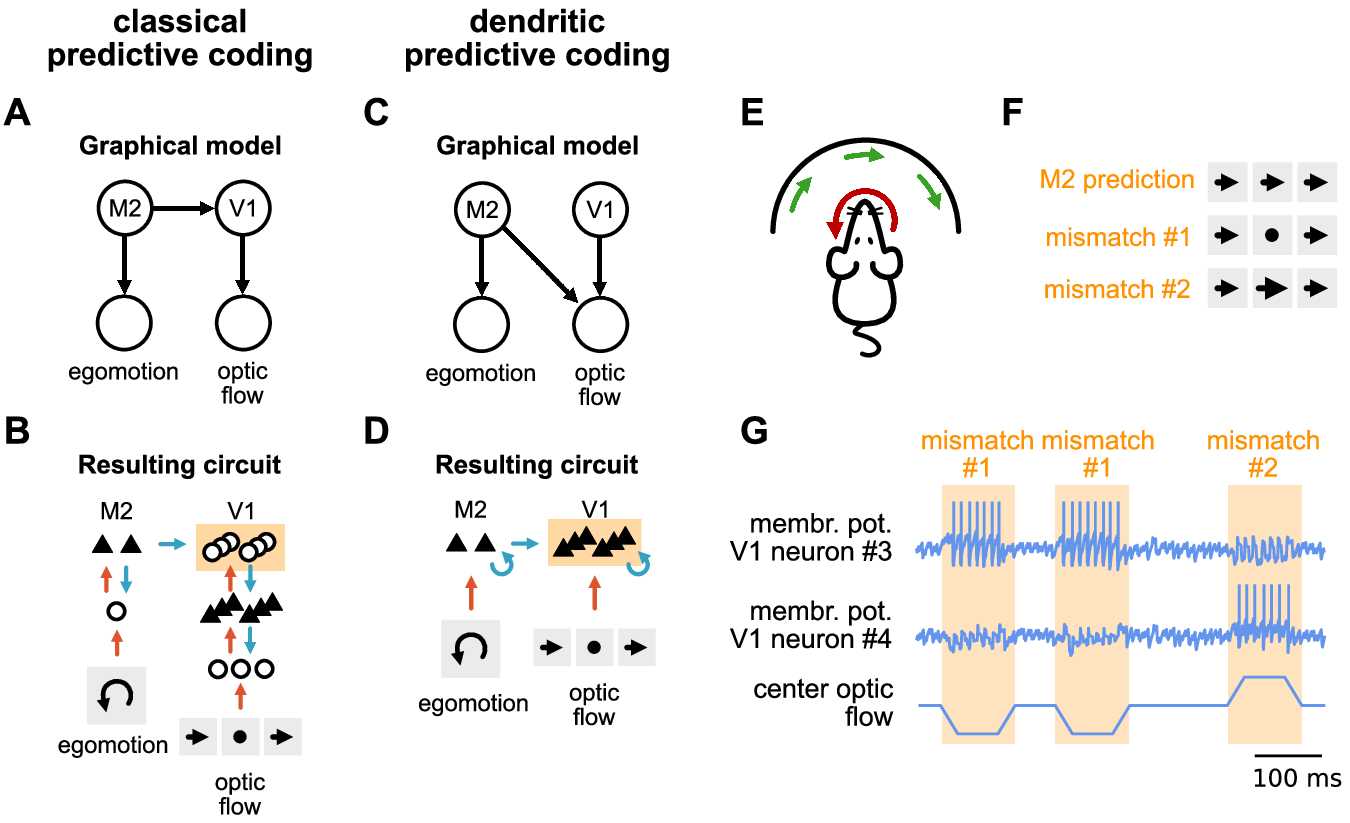}
    \caption{\textbf{Different models can explain optic flow mismatch responses for classical \hpc{} and \dec{}.} For a basic explanation of how to interpret graphical models, please refer to Infobox~\ref{info:generative_models}. 
    \textbf{A}~In classical \hpc{} we assume that motor-related neurons (M2) encode egomotion and visual neurons (V1) encode optic flow. One possible explanation for mismatch responses in V1 is that M2 also explains activity in V1. 
    \textbf{B}~In the resulting circuit this explanation of V1 by M2 introduces error neurons that would show mismatch responses (orange box). These neurons are driven by visual input, while M2 tries to predict (i.e balance) this input, resulting in neural responses when drive and prediction do not match~\cite{hertag2020learning}. 
    \textbf{C} In \dec{} M2 is seen as explaining optic flow jointly with V1. This introduces \textit{explaining away} effects between V1 and M2. 
    \textbf{D}~Explaining away means that in the corresponding neural circuit M2 should learn to balance dendrites of V1 neurons. This effectively cancels activity in V1 that can be predicted from M2. Hence, mismatch responses emerge in V1 prediction neurons (orange box) by the same superficial mechanism as in error units of classical \hpc{}. 
    \textbf{B,D}~For clarity, only connections that are essential for mismatch responses are shown. 
    \textbf{E}~Experimental setup to induce mismatch responses \cite{jordan2020opposing}. A mouse is placed in a virtual environment, while the head is fixed. Egomotion of the mouse (red arrow) results in visual flow (green arrows) that is displayed on a screen. 
    \textbf{F} Sample optic flow stimuli that are presented in our model. A rotation to the left would predict uniform visual flow to the right. Two mismatch conditions are also presented, where center optic flow is slower (\#1) or faster (\#2) than expected. 
    \textbf{G} Simulation of optic flow mismatch responses with the circuit in (D), reproduced from \cite{explainig_away_paper}. We find one visual neuron that learned to react specifically to slower center optic flow than expected (neuron \#3), and one that reacts to faster flow than expected (neuron \#4), which is similarly found in experiment~\cite{jordan2020opposing}.} 
    \label{fig:mismatch}
\end{figure}

A prime example of mismatch responses has recently been observed in visual cortex in neurons responding to optic flow~\cite{zmarz2016mismatch,jordan2020opposing}. In this specific experimental setup mice move through a virtual environment (Fig \ref{fig:mismatch}E,F). When a mismatch between optic flow predicted from egomotion and the presented optic flow is induced, this mismatch is signaled by V1 layer 2/3 neurons~\cite{jordan2020opposing}. The interpretation of these signals in terms of error neurons is straightforward: Predictions of optic flow from motor-related areas (M2) are subtracted from prediction neuron activity in V1 that encode \textit{all} optic flow, and the result of this subtraction is represented in layer 2/3 error neurons (Fig \ref{fig:mismatch}B).
In \dec{} the explanation of such mismatch responses is subtly different:
Here, motor and visual areas can be thought to jointly encode optic flow. Hence, V1 layer 2/3 prediction neurons should encode the \textit{residual} optic flow when self-generated optic flow is canceled via efference copies (Fig~\ref{fig:mismatch}C,D,G)~\cite{explainig_away_paper,aitchison_or_2017}. 
Such a representation is efficient, since here the redundancy between responses in V1 and M2 is reduced. It can also be very useful, for example to enable a moving observer to rapidly identify objects that move independently of a static surrounding~\cite{leinweber2017sensorimotor,guitchounts2020encoding,bouvier2020head}
(A common belief is that also mismatch responses of dedicated error neurons in classical \hpc{} could serve these two purposes, which we have argued is not the case; see \cite{explainig_away_paper} for details).
This interpretation of mismatch responses could explain why mismatch responses in V1 seem to be mixed with prediction responses~\cite{zmarz2016mismatch,fiser_experience-dependent_2016}.
Similar explanations can also be applied to other such observations, e.g. audio-visual suppression in V1~\cite{garner2021cortical}.
If correct, this could signify that cortex, already at the earliest levels of processing, encodes sensory stimuli as a whole in order to improve efficiency and integrate multisensory information.

\begin{figure}
\begin{infobox}[label={info:generative_models}]{Basic notions of generative models}
\begin{multicols}{2}
\small

\begin{minipage}[t]{0.485\textwidth}
\begin{center}
\includegraphics{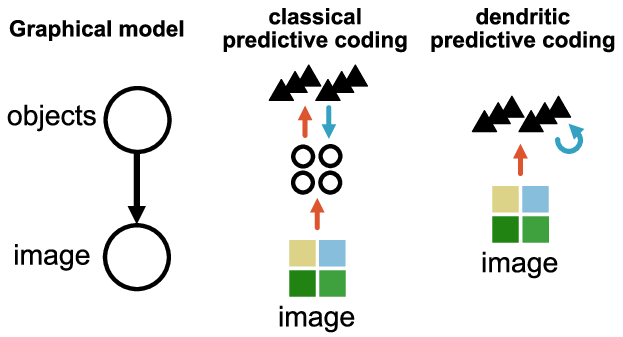}
\end{center}
\footnotesize\textit{Fig~B.} Example graphical model where images are explained by the presence of objects. Inference in this model can be performed with error neurons (small circles), as in classical \hpc{}, or with dendritic error computation, as in \dec{}. The resulting responses of prediction neurons (triangles) in principle are the same in both cases. In this example prediction neurons would be object detectors.
\end{minipage}
\vspace{1mm}

The mathematical foundation of \hpc{} are generative models, which are probabilistic models of (sensory) data. They formalize the idea that the brain builds a model of sensory data, and hence the world. The neural architectures, dynamics and plasticity rules always arise from a specific generative model~\cite{bastos_canonical_2012, rao_predictive_1999}, which is why it is helpful to understand the basic notions of generative modeling when talking about \hpc{}. 

A generative model assumes that the \textit{observed} distribution of sensory data can be well described by a process that generates the data from \textit{unobserved} (latent) variables. This process is formalized in terms of a conditional probability distribution $p_\theta(\text{data}|\text{latent})$, where $\theta$ are parameters of the distribution. In order to ease reasoning with these conditional probability distributions, the generative model is often depicted in the form of a graphical model, which expresses conditional dependencies between variables (Fig~B). In this example, it is assumed that images in a dataset can be generated by combining different objects (plus some noise), which is indicated by the arrow. 

In \hpc{} (or causal inference in general) it is assumed that, given the sensory data and prior expectations, the brain infers a consistent \textit{explanation} (also referred to as \textit{representation} or \textit{prediction}) of the data in terms of latent variables. This requires to \textit{invert} the generative model that the brain (implicitly) has of the world, i.e. to compute $p_\theta(\text{latent}|\text{data})$. This inference in the inverted model can be executed by a neural network, where the activity of prediction neurons corresponds to the value of certain latent variables (Fig~B). 

\textbf{Explaining away.} In many instances the data is not only explained by a single variable, but by several at once (Fig~C). Images, for example, could be explained in terms of objects and faces. In this case, when inferring the best explanations from the data, one latent variable must not explain aspects of the data that are already accounted for---i.e. \textit{explained away}---by other variables. In the networks this typically introduces inhibitory interactions between prediction neurons, as neurons should not encode what is already encoded by other neurons~\cite{moreno2015causal}. In classical \hpc{} this interaction is mediated by error neurons, while in \dec{} prediction neurons directly inhibit each other at the dendrites (Fig~C). Note, that in general explaining away already occurs between prediction neurons within a single level, as all these neurons explain the same data.

\vspace{5mm}
\begin{minipage}[t]{0.485\textwidth}
\begin{center}
\includegraphics{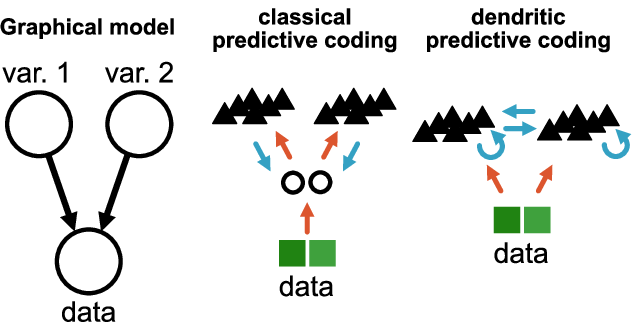}
\end{center}
\footnotesize\textit{Fig C.} In this generative model two variables explain the same data. Different neural interactions arise from this model in classical \hpc{} and \dec{}.
\end{minipage}

\end{multicols}
\end{infobox}
\end{figure}

Another central experimental observation that has been explained with error neurons in \hpc{} is the extra-classical receptive field effect of endstopping~\cite{rao_predictive_1999}. To induce this effect, bars with a fixed orientation but varying lengths are presented to a subject, and activity in visual cortex (layer 2/3 or layer 4) is measured. Two surprising observations can be made: First, the response of a neuron with a matching orientation tuning increases with increasing length when the bar is short, but when the bar extends over the classical receptive field, the response strongly decreases with increasing length~\cite{bolz1986generation}. Second, this extra-classical receptive field effect is reduced when feedback from higher-level areas is disabled~\cite{bolz1986generation}. 
In classical \hpc{} these effects are explained with a strong prior expectation of a higher-level area to observe long bars, which occur more frequently in natural images than short bars. Error neurons in a lower-level area will thus respond stronger for short bars, since they are less predicted from high-level feedback~\cite{rao_predictive_1999}. 
In \dec{}, where only prediction neurons are present, endstopping cannot be explained by this mechanism. However, recent theoretical work showed that endstopping behavior (and other extra-classical receptive field effects) also occurs in prediction neurons, where top-down connections strengthen this effect \cite{boutin_sparse_2021,spratling2010predictive}. Here, endstopping is mainly mediated by lateral inhibition between neurons with overlapping receptive fields \cite{spratling2010predictive}. Top-down connections from higher-level areas predict the activity patterns that arise from these lateral interactions and enhance them, which strengthens endstopping behavior~\cite{boutin_sparse_2021}.
This cooperation of lateral and top-down interactions could be important to help the network to cope with noise in the inputs and improve visual processing~\cite{boutin_sparse_2021}, and has been widely observed in visual cortex~\cite{liang2017interactions,nassi2013corticocortical,nurminen2018top,gilbert2013top,marques2018functional}.

\section{Open challenges for dendritic predictive coding}
As we have seen, the idea that neurons compute errors in their dendrites for hierarchical processing is consistent with many aspects of cortical connectivity and dynamics, but it also raises many questions. 
Here, we will discuss the most central open challenges that address, first, the micro-level of dendritic organization, second, the level of neural populations, and, third, the macro-level of how \dec{} could be implemented in the laminar organization of neocortex.

On the micro-level \dec{} employs a very simple picture of dendritic organization, whereas in reality neural dendrites are rather complex entities~\cite{bono2017modelling}. First, to be mathematically equivalent to classical \hpc{}, \dec{} assumes that single bottom-up connections and corresponding balancing lateral connections are perfectly separated from other inputs to the neuron \cite{mikulasch_2021}. Such a strict separation would require a large amount of balancing lateral connections,
and is not present in neural dendrites, where it is known that correlated excitatory synapses cluster together~\cite{kastellakis_synaptic_2015,chen_clustered_2012,kleindienst_activity-dependent_2011}. 
One suggestion is that synapses in a cluster could share predictions that are relevant for all these synapses, and by that minimize the number of lateral connections~\cite{mikulasch_2021}. In the beginning of learning, local inhibition could provide a backbone on which excitatory synapses are organized~\cite{kirchner2021emergence_review}. 
How exactly such a plasticity scheme could enable the locally coordinated learning of predictions is still unclear. 
Second, neural dendrites integrate synaptic inputs in a nonlinear fashion~\cite{goetz2021active}, which can greatly enhance the computational power of neurons~\cite{bono2017modelling}. The \dec{} model as we discuss it here, on the other hand, so far has only been derived under the assumption of linear dendritic integration for the linear encoding of signals. Future work will have to understand how nonlinear dendrites can be integrated into the framework, and how they can enable a nonlinear, and thus more versatile, encoding of sensory signals. 

On the level of neural populations it has to be understood how exactly \dec{} relates to cortical lateral connectivity.
In section \ref{sec:plausibility} we have already discussed the question of how the diversity of inhibitory interneurons supports the computations in \dec{}.
Another interesting issue is how the extensive excitatory lateral connectivity (which seems especially prevalent within layer 2/3) \cite{douglas_neuronal_2004} can be incorporated into the framework. Evidence suggests that cortex operates in a reverberating regime, which allows to integrate information over tuneable timescales and could be mediated by lateral excitatory connectivity~\cite{wilting2019between}. Similarly, it is thought that lateral connections provide a substrate for associative memory and pattern completion in cortex \cite{carrillo2020playing,osada2008towards}. Such mechanisms have been modeled in single-level models of spiking neurons \cite{kappel_stdp_2014,boerlin_predictive_2013,bill_distributed_2015}. Here, functional assemblies and temporal sequences can be learned with spike-timing dependent \cite{kappel_stdp_2014,bill_distributed_2015} or voltage-dependent plasticity \cite{clopath2010connectivity}.
An open question is how these mechanisms can purposefully interact with the learning of predictions in a hierarchical model (but see \cite{heeger2017theory} for a suggestion without learning).

\begin{figure}[!t]
    \centering
    \includegraphics{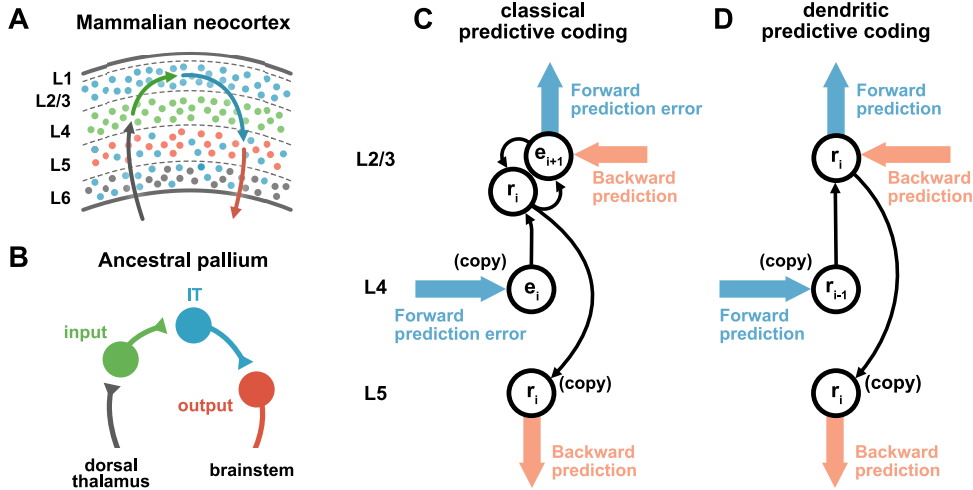}
    \caption{\textbf{How could dendritic predictive coding be embedded into neocortical microcircuits and lamination?} 
    \textbf{A}~Core circuitry of mammalian neocortex, as shown in~\cite{briscoe2018homology,dugas2012cell}. Input neurons in layer 4 (green) receive sensory information from the dorsal thalamus, layer 2/3 neurons further process this information, and output neurons in layer 5 (red) project to the brainstem.
    \textbf{B}~Theories of cortical evolution hypothesize that these input, intratelencephalic (IT) and output cells are homologous to cells that existed in the ancestral amniote pallium~\cite{briscoe2018homology}. In birds and non-avian reptiles these cells are organized in architectures that differ from the laminar organization of mammalian neocortex~\cite{briscoe2018homology}.
    \textbf{C}~The predictive coding microcircuit as proposed by \cite{bastos_canonical_2012} (here presented in a simplified form) follows the organization of the neocortical microcircuit. Predictions ($r_i$) and prediction errors ($e_i$) are computed in layer 2/3. Deeper layers mainly act as communication hubs by copying signals from layer 2/3. 
    \textbf{D}~Speculative microcircuit for dendritic predictive coding. Here, deeper layers fulfill the same function as communication hubs, but layer 2/3 only computes predictions. More connections (especially within layer 2/3 between areas) might exist \cite{harris2015neocortical}, but are left out here for simplicity.}
    \label{fig:microcircuit}
\end{figure}

On the macro-level of cortical circuits and lamination, an important question is what the computational roles of pyramidal neurons in the different layers of (neo-)cortex are~\cite{harris2015neocortical}. In classical \hpc{}, based on neocortical connectivity, an influential theory of these roles has been brought forward by Bastos et al~\cite{bastos_canonical_2012} (Fig~\ref{fig:microcircuit}C). This theory states that errors and predictions are first computed in layer 2/3, while layer 4 and 5 neurons primarily redistribute signals. 
This role of deeper cortical layers is also compatible with theories of cortical evolution~\cite{shepherd2017neocortical}. Ancestral amniote (reptiles and mammals) cortex was based on a 3-layer architecture, with a single layer of pyramidal neurons, and is thought to have operated relatively independently of other brain structures. Deeper layers in neocortex might have migrated from previously separate neural populations to cortex in order to integrate cortical neurons more deeply with the rest of the brain and other cortical areas~\cite{shepherd2017neocortical}. This idea is supported by evidence from studies of brain anatomy and cell-type markers~\cite{dugas2012cell,karten2013neocortical,briscoe2018homology} (Fig~\ref{fig:microcircuit}A,B). 

For \dec{} we therefore suggest the working hypothesis that, similar to the theories of Bastos et al~\cite{bastos_canonical_2012} and Douglas and Martin~\cite{douglas_neuronal_2004}, deeper layers mainly act as long-range communication hubs for cortical integration, while the associative and computational role of the neurons in the \dec{} model is performed by pyramidal neurons in layer 2/3 (Fig~\ref{fig:microcircuit}D).
Cortical connectivity can support this idea: Layer 2/3 pyramidal neurons would send predictions, directly or mediated by layer 4 neurons, up the hierarchy where they are integrated at basal dendrites, and, directly or mediated by layer 5 neurons, down the hierarchy where they are integrated at apical dendrites~\cite{douglas_neuronal_2004,bastos_canonical_2012,harris2015neocortical}. 
Also the different lamina-specific coding properties of pyramidal cells are compatible: Layer 2/3 neurons exhibit sparse activity, which is suited for the computational role of \dec{}~\cite{avermann2012microcircuits,petersen2013synaptic,harris2013cortical}. As discussed before, the Gamma rhythm of these neurons could be indicative of an efficient neural code. Layer 5 neurons, in contrast, employ a dense code, which can effectively broadcast information to long-range targets~\cite{harris2013cortical}.
Future work will have to revisit this working hypothesis, and examine in what sense the suggested role of deeper layers is an oversimplification (possible, complementary roles have been discussed before, e.g.~\cite{douglas_neuronal_2004}). 

\section{Summary \& Outlook}

Since its conception over twenty years ago, hierarchical predictive coding has been considered one of the most promising unifying theories of cortical computation, but---in its classical form---it is still facing substantial questions regarding its biological plausibility. 
Here, we outlined an emerging hierarchical predictive coding scheme based on dendritic error computation, which is functionally equivalent, but provides solutions to the most glaring open problems of the established theory of classical \hpc{}: First, it can explain the lack of clear empirical evidence for the coexistence of error and prediction neurons~\cite{walsh2020evaluating,heilbron_2018}, and second, it overcomes the unresolved question of how learning can be efficiently implemented with spiking error neurons~\cite{millidge_predictive_2021}.
Moreover, it has the potential to explain a broad range of additional experimental observations:
From the idea that prediction errors are represented at neural dendrites, it directly follows that synaptic inputs to basal dendrites should be locally balanced by lateral inhibition, which is a network motif that is indeed ubiquitous in cortex~\cite{douglas_neuronal_2004,deneve_efficient_2016,iascone2020whole,gillon2021learning}. 
Also, the learning rules derived in \dec{} are consistent with the observation that synaptic plasticity is strongly modulated by the local voltage in dendrites~\cite{hayama2013gaba,clopath2010voltage}.
Furthermore, \dec{} can relate these microscopic properties of neural dendrites to neural dynamics and learning on macroscopic scales, and shows how they could give rise to the neural spike response characteristics~\cite{deneve_efficient_2016}, receptive fields~\cite{mikulasch_2021,boutin_sparse_2021} and the influence of top-down connections~\cite{boutin_sparse_2021}, that are observed in cortex. 

These advances open up several exciting paths for future research. 
Empirically, the predictions of \dec{} should be tested directly in experiments, and specialized experiments should be designed that can distinguish between the two competing theories of classical and dendritic \hpc{}. 
The latter will be an auspicious but challenging endeavor: 
First, because classical \hpc{} as of yet does not make clear predictions on the level of single neurons. 
Second, because often indirect measures of neural activity (e.g. EEG, fMRI) have been used to search for evidence of classical \hpc{}, and it remains to be worked out in detail how error computation in specialized neurons and in dendrites would differ in these measures. 
Third, and most importantly, because the expected computations of prediction neurons are the same in the two theories---even though classical and dendritic \hpc{} give competing answers to the question of how exactly inference is performed in neural circuits, they still are just two possible implementations of the same underlying mathematical framework. 
This, however, also creates opportunities: Many of the important conceptual insights that have been gained over the last years in classical \hpc{}, such as what role the precision of predictions could play in cortical computations~\cite{friston2018does}, can be transferred to \dec{}, where they will acquire new meaning in terms of neural physiology and dynamics.



\section*{Acknowledgements}
We want to thank Abdullah Makkeh, Beatriz Belbut, Caspar Schwiedrzik, Georg Keller and the Priesemann Lab, especially Andreas Schneider and Matthias Loidolt, for helpful discussions and comments on the manuscript.

\printbibliography

\end{document}